\documentclass[11pt]{article}

\usepackage{acl}

\usepackage{times}
\usepackage{latexsym}
\usepackage[T1]{fontenc}
\usepackage[utf8]{inputenc}

\usepackage{microtype}

\usepackage{inconsolata}

\usepackage{graphicx}

\usepackage{listings}

\usepackage{algorithm}
\usepackage{algorithmicx}
\usepackage{algpseudocode}

\usepackage{comment}
\usepackage{url}

\algrenewcommand\algorithmiccomment[1]{%
  \hfill$\triangleright$~\parbox[t]{0.55\linewidth}{#1}%
}

\lstdefinestyle{promptstyle}{
    basicstyle=\footnotesize\ttfamily,
    frame=single,
    breaklines=true,
    breakatwhitespace=true,
    columns=fullflexible,
    aboveskip=2pt,
    belowskip=2pt,
    framesep=3pt,
    xleftmargin=0pt,
    xrightmargin=0pt,
}

\usepackage{booktabs}
\usepackage{siunitx}
\usepackage{multirow}

\usepackage{amsfonts}

\usepackage{tikz}
\usetikzlibrary{arrows.meta,positioning,fit,calc}

\title{\textsc{StructSurvey}: Structured Agentic Retrieval \\ for Automated Survey Paper Generation}

\author{
Paolo Pedinotti\thanks{Paolo Pedinotti contributed to this work during his internship at Bloomberg.}\\
Bloomberg\\
\texttt{pedinotti.paolo@gmail.com}
\And
Enrico Santus\\
Bloomberg\\
\texttt{esantus@bloomberg.net}
}

\begin{document}
\maketitle

\begin{abstract}
The rapid growth of scientific publications makes it increasingly difficult to track and synthesize research progress. While Large Language Models (LLMs) can support automated survey generation, existing methods retrieve unstructured data and require models to infer conceptual, methodological, and taxonomic relations from raw text at generation time. We introduce \textsc{StructSurvey}, a hierarchical multi-agent framework that shifts structural reasoning from generation to retrieval by dynamically constructing graph-based representations of entities, relations, and topical taxonomies. We evaluate \textsc{StructSurvey} on a new reference-grounded benchmark of ACL survey papers for reproducible long-form scientific summarization. Compared with embedding-only retrieval baselines, \textsc{StructSurvey} improves ROUGE-1 recall by +2.9 and ROUGE-2 recall by +1.0 on average, without reducing precision. It also improves LLM-as-a-Judge ratings for logical structure, depth, and synthesis, showing that explicit structural retrieval yields surveys closer to human-written organization and reasoning.
\end{abstract}

\section{Introduction}

\begin{figure*}[t]
\centering
\includegraphics[width=\textwidth]{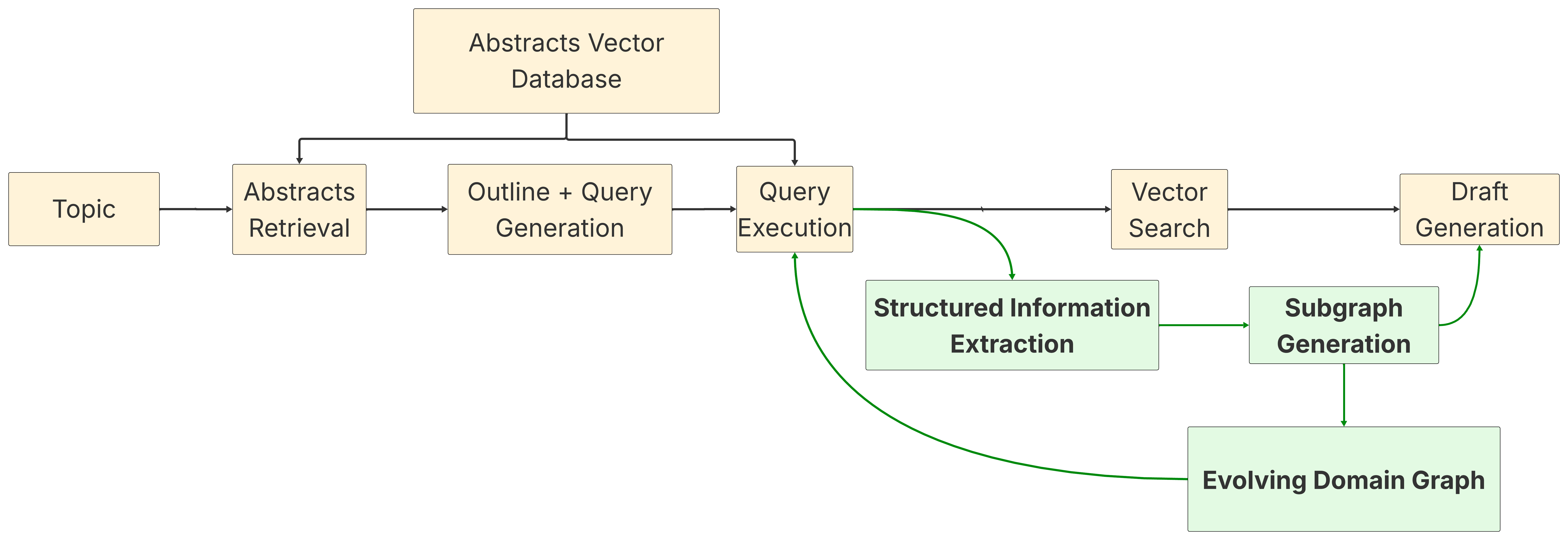}
\caption{\textsc{StructSurvey} process. Unlike vector-only retrieval systems (yellow), \textsc{StructSurvey} extracts structured information (green), merges retrieved subgraphs into an evolving domain graph, and uses it to guide generation.}
\label{fig:system_illustration}
\end{figure*}

The rapid growth of scientific publications has made it increasingly difficult for researchers to identify, interpret, and synthesize relevant developments. Scientific surveys address this challenge by organizing fragmented contributions into coherent narratives that reveal conceptual structure, methodological evolution, and emerging trends. However, writing high-quality surveys remains labor-intensive, requiring domain expertise, careful source selection, and substantial time.

This has motivated work on \emph{automated survey paper generation} \cite{NEURIPS2024_d07a9fc7, yan-etal-2025-surveyforge}, where Large Language Models (LLMs) assist in retrieving, organizing, and synthesizing research literature. Survey generation is particularly demanding because it requires not only factual grounding, but also the ability to identify relationships among methods, tasks, datasets, and research directions across many papers.

Existing systems typically rely on \textbf{Retrieval-Augmented Generation (RAG)} \cite{10.5555/3495724.3496517}, retrieving semantically relevant documents and passing them to the LLM as evidence. While this improves grounding, current approaches use \emph{unstructured retrieval}: retrieved passages provide no explicit representation of how ideas relate, how entities cluster, or how methods connect across the literature. This is a key limitation for survey writing, where quality depends not only on retrieving relevant papers but also on organizing them into coherent conceptual and methodological groupings. As a result, LLMs must reconstruct the conceptual scaffolding of a survey implicitly during generation, which is especially difficult for hierarchical documents organized into sections, subsections, and fine-grained themes.

We introduce \textsc{StructSurvey}, a hierarchical multi-agent framework that incorporates \emph{structured retrieval} \cite{Jiang_2025} into survey generation. Instead of relying only on vector search, \textsc{StructSurvey} uses \textbf{parameterized query functions} that trigger on-demand extraction of entities, relations, and taxonomic groupings from retrieved abstracts. These outputs are merged into an evolving domain graph that captures relationships among methods, tasks, and research directions as the outline expands. Figure~\ref{fig:system_illustration} illustrates this process. By shifting structural reasoning from generation to retrieval, \textsc{StructSurvey} provides explicit, interpretable context for outline formation and narrative synthesis.

To evaluate this approach, we construct a new \textbf{ACL Survey Dataset} of 33 survey papers published between 2018 and 2025, together with their full text and publicly accessible referenced papers. This reference-grounded benchmark enables reproducible comparison of survey-generation systems under controlled evidence conditions.

On this benchmark, \textsc{StructSurvey} improves over embedding-only retrieval baselines by +2.9 ROUGE-1 recall and +1.0 ROUGE-2 recall on average, while preserving comparable precision. These gains show that structured retrieval provides more relevant evidence during outline expansion and drafting. Since lexical overlap cannot fully capture survey quality, we also conduct \textbf{LLM-as-a-Judge evaluations} \cite{gu2025surveyllmasajudge} using a robustness-oriented protocol over logical structure, depth, critical analysis, and synthesis. \textsc{StructSurvey} improves all dimensions except critical analysis, which remains a limitation across systems.

Our contributions are:
\begin{itemize}
    \item \textbf{\textsc{StructSurvey}}, a hierarchical multi-agent framework that uses structured retrieval and evolving domain graphs to guide survey planning, retrieval, and writing;
    \item a \emph{reference-grounded} benchmark comprising 33 ACL survey papers published between 2018 and 2025, all with publicly accessible references (see Appendix~\ref{app:dataset_papers});
    \item a reproducible \textbf{LLM-as-a-Judge evaluation protocol} designed to reduce prompt sensitivity, model bias, and reproducibility issues;
    \item empirical evidence that structured retrieval improves both lexical overlap and higher-level survey qualities such as logical structure, depth, and synthesis.
\end{itemize}

\section{Related Work}

Automated survey paper generation with LLMs is an emerging research area. Recent systems generally frame survey generation as a hierarchical planning and writing task, where LLM agents retrieve relevant literature, generate outlines, and draft sections. The most directly comparable recent systems are \textsc{AutoSurvey} \citep{NEURIPS2024_d07a9fc7}, which introduced a systematic LLM-based pipeline for long-form survey writing, and \textsc{SurveyForge} \citep{yan-etal-2025-surveyforge}, which extends this line of work with heuristic outline generation, memory-driven retrieval, and multi-dimensional evaluation.

\paragraph{AutoSurvey.}
\textsc{AutoSurvey} formulates survey generation as a multi-stage workflow consisting of initial retrieval and outline generation, subsection drafting, integration and refinement, and final evaluation. The system first retrieves papers relevant to the survey topic, generates and merges candidate outlines, and then drafts sections in parallel using subsection-specific retrieval. It further refines generated sections for coherence and citation correctness, and uses LLM-based evaluation to select among generated survey candidates. However, retrieval is based on embedding similarity, and the retrieved evidence is passed to downstream generation modules as unstructured textual context. As a result, reasoning about how concepts, methods, and themes relate is largely left to the generation phase, where the LLM must organize long and heterogeneous evidence without an explicit representation of conceptual or taxonomic structure.

\paragraph{\textsc{SurveyForge}.}
\textsc{SurveyForge} adopts a related hierarchical generation architecture but introduces two important extensions. First, it uses both a research-paper database and a survey-outline database, allowing human-written survey structures to guide outline generation. Second, it expands outlines recursively and employs a memory-driven Scholar Navigation Agent with sub-query decomposition and temporally-aware reranking to retrieve higher-quality references for each subsection. Despite these additions, \textsc{SurveyForge} still builds primarily on embedding-based retrieval, augmented with memory and reranking, and provides downstream agents with textual evidence rather than explicit relational or taxonomic representations.

\paragraph{Graph-based Retrieval.}
Recent work on graph-based retrieval-augmented generation (GraphRAG) investigates the use of graph-structured knowledge for reasoning and generation \citep{peng2024graphretrievalaugmentedgenerationsurvey, Han2024RetrievalAugmentedGW, edge2024fromlocaltoglobal}. These approaches often construct or assume a graph representation over a corpus, such as a knowledge graph, citation network, or entity-relation graph. However, survey writing requires topic-specific and outline-dependent structure. %\textsc{StructSurvey} therefore constructs and expands a domain graph dynamically during generation. At each outline-expansion step, LLM agents extract local subgraphs of entities, relations, and taxonomic groupings, then merge them into an evolving knowledge structure that conditions later retrieval and writing.

\paragraph{Comparison.}
Existing survey-generation systems largely treat the task as a tree-structured pipeline over retrieved textual evidence. Consequently, the LLM must infer relationships among concepts, methods, datasets, tasks, and research threads during generation. \textsc{StructSurvey} addresses this limitation by shifting structural inference into retrieval: query functions can return explicit knowledge representations, including extracted entities, relations, and lightweight taxonomies derived from retrieved abstracts. Unlike static GraphRAG settings, this structure is created on demand for each outline node and reused as the survey develops. This enables more targeted retrieval and more coherent generation, while remaining model-agnostic and compatible with previous frameworks.

\section{Method Overview}
\label{sec:method}

The core insight behind \textsc{StructSurvey} is that survey papers are inherently structured artifacts: they organize a research area into entities (e.g., models, datasets, tasks), relations among those entities, and higher-level taxonomic groupings. Instead of requiring an LLM to reconstruct this structure implicitly from unstructured text during generation, \textsc{StructSurvey} explicitly extracts, accumulates, and reuses structured knowledge throughout the survey-generation process.

\paragraph{Workflow Overview.}
The \textsc{StructSurvey} workflow consists of three stages that are applied hierarchically as the survey outline is expanded:

\begin{enumerate}
    \item \textbf{Outline planning.}
    Given a survey topic, the system retrieves an initial set of candidate papers using embedding-based retrieval over a paper corpus. A planner LLM generates a first-level outline, where each section is paired with a \emph{parameterized query function}. The query function specifies how evidence for that section should be retrieved and organized.

    \item \textbf{Structured retrieval and graph construction.}
    Executing a query function triggers either (i) standard vector retrieval, which returns a set of relevant abstracts, or (ii) structured extraction, which produces a \textbf{local graph fragment} encoding entities, relations, or taxonomic assignments extracted from the papers. These local graph fragments are incrementally merged into an evolving domain graph that accumulates structured knowledge across outline levels.

    \item \textbf{Section drafting.}
    Writer LLMs decompose each section-level query into finer-grained subqueries, retrieve both unstructured text and structured graph context, and draft subsections in parallel. Subsection drafts are then merged to form complete section drafts, which are finally assembled into the full survey.
\end{enumerate}

This workflow operationalizes survey generation as a repeated sequence of \emph{planning $\rightarrow$ structuring $\rightarrow$ writing}, allowing LLMs to condition generation on explicit, interpretable structure rather than raw text alone.

\paragraph{Query Interface.}
The central abstraction in \textsc{StructSurvey} is the \textbf{query function}. A query function is a typed retrieval instruction attached to each outline node. Unlike prior systems, which associate sections only with free-text queries, \textsc{StructSurvey} uses query functions to explicitly control both the retrieval strategy and the structure of the returned evidence.

Each query function returns either:
(i) a set of retrieved texts (vector retrieval), or
(ii) a structured, undirected graph fragment extracted from the texts (structured retrieval).

During outline generation, the planner LLM acts as a \textbf{routing agent}, selecting the best query function for each section.

The query functions implemented in \textsc{StructSurvey} are:

\begin{itemize}
    \item \texttt{single\_entity\_search}.
    This function extracts all entities of a specified type and organizes them according to a categorization criterion.
    \textbf{Example:} extracting datasets mentioned in the abstracts and grouping them by task type (e.g., classification, generation, parsing).
    \textbf{Output:} a graph whose nodes correspond to datasets, annotated with category labels and frequency counts.

    \item \texttt{related\_entity\_search}.
    This function extracts entities that are explicitly related to a fixed entity.
    \textbf{Example:} extracting downstream NLP tasks associated with the BERT model and categorizing them by linguistic level.
    \textbf{Output:} a graph centered on the fixed entity (e.g., BERT), with edges to related entities.

    \item \texttt{pair\_of\_related\_entities\_search}.
    This function extracts pairs of related entities and assigns each side to its own taxonomy.
    \textbf{Example:} extracting transformer architectures and the NLG tasks they are applied to, yielding pairs such as (\textit{BART}, summarization), (\textit{T5}, data-to-text generation), and (\textit{mBART}, machine translation).
    \textbf{Output:} a bipartite graph linking source and target entities, with category annotations on both sides.
\end{itemize}

\paragraph{Structured Extraction Utilities.}
Each query function is implemented as a composition of LLM-based extraction utilities. Following \citet{edge2024fromlocaltoglobal} and \citet{wan2024tntllmtextminingscale}, \textsc{StructSurvey} employs the following components:

\begin{itemize}
    \item \textbf{Entity extraction}: parallel extraction of entities across documents (Prompt~\ref{sec:entity_extraction_prompt}).
    \item \textbf{Relationship-based extraction}: identification of entities related to a specified fixed entity (Prompt~\ref{sec:relationship_based_entity_extraction_prompt}).
    \item \textbf{Entity pair extraction}: extraction of typed source--target entity pairs (Prompt~\ref{sec:entity_pair_extraction_prompt}).
    \item \textbf{Category discovery}: induction of a compact taxonomy (3-8 categories) covering a set of extracted entities (Prompt~\ref{sec:category_discovery_prompt}).
    \item \textbf{Entity classification}: parallel assignment of extracted entities to the discovered categories (Prompt~\ref{sec:entity_classification_prompt}).
\end{itemize}

\paragraph{Graph Construction and Merging.}
Each structured query produces a local undirected graph fragment corresponding to a specific outline node. These fragments are merged into a global domain graph by canonicalizing entity strings, merging nodes with identical canonical forms, uniting edges and category assignments, and aggregating frequency statistics across extractions. The resulting domain graph grows incrementally as the outline deepens and is reused across subsequent retrieval and generation steps.

\paragraph{Graph Conditioning of Generation.}
When drafting sections, the global domain graph or the relevant subgraph is serialized into a compact textual representation consisting of entity lists grouped by category, explicit entity--entity relations, and frequency statistics. This serialized graph context is injected directly into the second-level outline generation and the writing prompts (Prompt~\ref{sec:second_level_outline_generation_prompt} and Prompt~\ref{sec:section_writing_prompt}), where the LLM is explicitly instructed to use it to organize and structure the generated narrative.

\paragraph{Algorithmic Implementation.}
Algorithm~\ref{alg:generate_survey} summarizes the full pipeline, combining outline planners, routing agents, query functions, extraction utilities, and writer agents into a coherent end-to-end system.

\begin{algorithm*}[t]
\small
\caption{\textsc{StructSurvey}: Survey Generation with Structured Retrieval and Graph Reuse}
\label{alg:generate_survey}
\begin{algorithmic}[1]

\Require Survey topic $T$; paper vector index $P$; query-function library $\mathcal{Q}$; prompt set $\Pi$
\Ensure Generated survey text $S$

\State Initialize global domain graph $G \gets \emptyset$
\State Initialize section drafts $\mathcal{D} \gets \emptyset$
\Comment{\textbf{Stage 1: Seed retrieval and first-level planning}}
\State $D_{\mathrm{top}} \gets \textsc{VectorSearch}(T, P)$ \Comment{Retrieve top-$k$ papers for topic $T$}
\State $\mathcal{O}^{(1)} \gets \textsc{LLM}(\Pi_{\mathrm{Outline}}^{(1)}, T, D_{\mathrm{top}}, \mathcal{Q})$ \Comment{Generate first-level outline with query functions}
\For{$i = 1$ to $N^{(1)}$}
    \Comment{\textbf{Stage 2: First-level retrieval and second-level planning}}
    \State $(O_i^{(1)}, Q_i^{(1)}) \gets \mathcal{O}^{(1)}_i$ \Comment{Unpack section title and assigned query}
    \State $(C_i^{(1)}, G_i^{(1)}) \gets \textsc{Execute}(Q_i^{(1)}, P, G)$ \Comment{Run vector or structured retrieval}
    \State $G \gets \textsc{MergeGraph}(G, G_i^{(1)})$ \Comment{Reuse and update global graph}
    \State $G_i^{\mathrm{ctx}} \gets \textsc{SelectSubgraph}(G, O_i^{(1)}, C_i^{(1)})$
    \State $\mathcal{O}_i^{(2)} \gets \textsc{LLM}(\Pi_{\mathrm{Outline}}^{(2)}, O_i^{(1)}, C_i^{(1)}, G_i^{\mathrm{ctx}})$ \Comment{Generate subsections}

    \Comment{\textbf{Stage 3--4: Sub-querying and drafting}}
    \State $\mathcal{D}_i \gets \emptyset$
    \For{$j = 1$ to $N_i^{(2)}$}
        \State $\mathcal{R}_{i,j} \gets \textsc{LLM}(\Pi_{\mathrm{Decompose}}, O_{i,j}^{(2)}, \mathcal{Q})$ \Comment{Create 3--5 retrieval sub-queries}
        \State $C_{i,j} \gets \emptyset$
        \For{each $Q_{\mathrm{sub}} \in \mathcal{R}_{i,j}$}
            \State $(C_{\mathrm{sub}}, G_{\mathrm{sub}}) \gets \textsc{Execute}(Q_{\mathrm{sub}}, P, G)$
            \State $C_{i,j} \gets C_{i,j} \cup C_{\mathrm{sub}}$
            \State $G \gets \textsc{MergeGraph}(G, G_{\mathrm{sub}})$ \Comment{Accumulate evidence and graph structure}
        \EndFor
        \State $G_{i,j}^{\mathrm{ctx}} \gets \textsc{SelectSubgraph}(G, O_{i,j}^{(2)}, C_{i,j})$
        \State $D_{i,j} \gets \textsc{LLM}(\Pi_{\mathrm{Write}}, O_{i,j}^{(2)}, C_{i,j}, G_{i,j}^{\mathrm{ctx}})$ \Comment{Draft subsection}
        \State $\mathcal{D}_i \gets \mathcal{D}_i \cup \{D_{i,j}\}$
    \EndFor
    \State $D_i \gets \textsc{MergeDrafts}(\mathcal{D}_i)$ \Comment{Merge subsection drafts}
    \State $\mathcal{D} \gets \mathcal{D} \cup \{D_i\}$
\EndFor
\Comment{\textbf{Stage 5: Final assembly}}
\State $S \gets \textsc{MergeDrafts}(\mathcal{D})$ \Comment{Assemble full survey}
\State \Return $S$

\end{algorithmic}
\end{algorithm*}

%\subsection{System Cost}

%We analyze system cost in terms of LLM call counts, which dominate both runtime and monetary cost in multi-agent pipelines. In \textsc{StructSurvey}, cost depends on the types of query functions executed during survey generation.

%\paragraph{Vector-based queries.}
%Vector-based queries rely on embedding similarity search over the paper index and incur no additional LLM calls beyond embedding computation. They typically run in $O(\log D)$ time on the index, where $D$ is the number of documents.

%\paragraph{Structured queries.}
%Structured queries rely on LLM-based information extraction and are more expensive. Let $E$ denote the number of extracted entities and $b$ the batch size for entity classification. A single structured query performs:
%\begin{itemize}
%    \item $D$ parallel LLM calls for entity extraction,
%    \item one LLM call for category discovery,
%    \item $\lceil E / b \rceil$ parallel LLM calls for entity classification.
%\end{itemize}
%The resulting LLM-call cost is
%\[
%    \mathrm{Cost}_{\mathrm{LLM}} = D + 1 + \left\lceil \frac{E}{b} \right\rceil .
%\]

%\paragraph{Reuse and amortization.}
%Structured queries yield local graph fragments that are merged into a global domain graph and reused across subsequent outline expansion and drafting steps (Algorithm~\ref{alg:generate_survey}). This reuse amortizes extraction cost across sections and reduces redundant computation. Overall cost scales linearly with the number of structured queries and the choice of LLM model, while vector-based retrieval remains comparatively inexpensive.

\section{Experiments}

\subsection{Data}
\label{sec:data}

To evaluate \textsc{StructSurvey} in a realistic scientific writing setting, we construct a gold-standard dataset of 33 ACL Anthology surveys published between 2018 and 2025, listed in Appendix~\ref{app:dataset_papers}.
Each survey is paired with the set of papers it cites, forming a benchmark for \emph{reference-grounded} evaluation. Under this setting, systems are evaluated only on content drawn from the same reference pool available to the human authors, preventing the use of external or future information and enabling fair comparison to human-written surveys.

\paragraph{Survey paper extraction.}
We queried the ACL Anthology to identify survey papers published between 2018 and 2025 in major ACL venues. Candidate surveys were identified using keywords indicative of surveys. 
%including \emph{survey}, \emph{systematic review}, \emph{overview}, and \emph{a review of}. 
This process yielded an initial pool of survey candidates spanning a broad range of NLP subfields.

\paragraph{Referenced paper collection.}
For each survey, we retrieved the complete reference list via Semantic Scholar.\footnote{\url{https://www.semanticscholar.org/product/api}} 
%including metadata for all cited papers
To replicate the conditions under which the original surveys were written, we retained only surveys for which at least 70\% of referenced papers were publicly available through the ACL Anthology or arXiv. This constraint avoids licensing issues and yields a clean, reproducible benchmark.

\paragraph{Dataset characteristics.}
Each dataset entry includes the survey abstract, full-body text (extracted using the Mistral OCR model\footnote{\url{https://mistral.ai/news/mistral-ocr}}), and the set of accessible referenced papers (with abstracts). Key corpus statistics are reported in Table~\ref{tab:corpus_statistics}.

\subsection{Experimental Setting}

We compare \textsc{StructSurvey} directly with \textsc{SurveyForge} under a controlled experimental setup. Both systems share the same multi-agent hierarchical architecture, maximum outline depth, and writing procedures. The only difference is the retrieval mechanism: \textsc{SurveyForge} supports only vector-based queries, whereas \textsc{StructSurvey} additionally allows LLMs to invoke structured query functions. This isolates the contribution of structured retrieval.

\paragraph{Model configuration.}
All LLM components run on \texttt{Gemini-2.5-Flash~Lite}, accessed via the official Gemini API.\footnote{\url{https://ai.google.dev/gemini-api/docs}} Embeddings for vector retrieval are generated with OpenAI \texttt{text-embedding-3-small}\footnote{\url{https://platform.openai.com/docs/models/text-embedding-3-small}} and indexed using FAISS \citep{douze2025faisslibrary}. 

The decoding configuration is kept identical across both systems: temperature $1$ and top-$p$ $0.95$.

\paragraph{Abstract-level retrieval.}
For both vector-based and structured retrieval, we operate exclusively on paper abstracts rather than full texts. This choice reflects a trade-off between computational cost, scalability, and experimental control: abstract-level retrieval enables processing across thousands of cited papers while avoiding the substantial cost associated with long-document extraction. Importantly, both \textsc{StructSurvey} and the baseline system are subject to the same constraint, ensuring a fair comparison. 
%While full-text retrieval could provide richer evidence, it would also introduce additional confounding factors related to document segmentation and context length, which we leave to future work.

\paragraph{First-level outline generation.}
The system retrieves the top 20 abstracts using embedding-based similarity with the survey topic. Using Prompt~\ref{sec:first_level_outline_generation_prompt}, the planner LLM produces 4-6 top-level sections, each paired with a query function. This pairing is the key decision point: \textsc{SurveyForge} always assigns a vector retrieval function, whereas \textsc{StructSurvey} may assign either a vector or a structured query function depending on the semantic cues in the retrieved context.

\paragraph{Second-level outline generation.}
For each first-level section:
\begin{itemize}
    \item If the query function is vector-based, the system retrieves the top 15 abstracts and generates 3-5 subsections using Prompt~\ref{sec:second_level_outline_generation_prompt}.
    \item If the query is structured, the system executes the corresponding structured extraction functions (described in Section~\ref{sec:method}) on the abstracts. The resulting structured graph is provided directly to the outlining LLM.
\end{itemize}
This stage expands each coarse section into a more fine-grained and topic-aware outline.

\paragraph{Content generation.}
After the hierarchical outline is finalized, writer LLMs decompose each section query into subqueries using Prompt~\ref{sec:query_decomposition_prompt}. The system retrieves evidence — either vector-based or structured — and passes the merged content to Prompt~\ref{sec:section_writing_prompt} to produce 500-800 word drafts. Prompts for query decomposition (except for the list of available query functions) and content generation are identical across systems, ensuring a fair comparison.

\paragraph{Reference-grounded evaluation.}
Our experimental setup evaluates survey-generation systems under a reference-grounded setting, where the retrieval corpus for each survey consists exclusively of the papers cited by the gold survey available on the ACL Anthology or arXiv. This design choice isolates the core challenge addressed in this work — planning, structuring, and synthesizing a research area given relevant literature — while controlling for variability introduced by open-ended document discovery. By constraining systems to the same evidence available to the human authors, we enable fair and reproducible comparison of generated surveys against gold references. At the same time, this setting abstracts away retrieval precision errors that arise in open-corpus scenarios; incorporating citation-graph expansion or open-domain retrieval is an important direction for future work but is orthogonal to the structured-retrieval mechanisms studied here.

\subsection{Evaluation}

Evaluating automatically generated surveys poses unique challenges: surveys are long, require domain-level knowledge, and must be assessed along multiple qualitative dimensions. Prior work has relied primarily on LLM-as-a-Judge assessment, but such evaluation introduces reliability concerns due to prompt sensitivity, model-specific biases, and limited reproducibility \citep{gu2025surveyllmasajudge, li2024llmsasjudgescomprehensivesurveyllmbased}. We therefore combine standard lexical overlap metrics with a principled LLM-guided evaluation protocol.

\paragraph{Metrics.}
We compute ROUGE-1 and ROUGE-2 precision, recall, and F1 between system-generated surveys and the corresponding gold surveys. Although ROUGE does not capture global structure or deep semantics, it provides a stable and reproducible measure of lexical and phrasal overlap that is sensitive to topical coverage (recall) and hallucination (precision).

\paragraph{LLM-guided evaluation.}
To assess higher-level qualities not captured by ROUGE, we complement lexical metrics with LLM-as-a-Judge evaluations. Following recent analyses of evaluation robustness \citep{baumann2025largelanguagemodelhacking}, we adopt a principled protocol that mitigates prompt sensitivity and model-specific biases.

First, we selected a pair of evaluator models, including both proprietary and open models. For our experiments, we use \texttt{claude-sonnet-4-5-20250929} and \texttt{deepseek.r1-v1:0} via Amazon Bedrock.\footnote{\url{https://aws.amazon.com/bedrock/}}

We then defined a set of criteria for survey quality, including outline-level criteria (involving the analysis of outlines), and full survey-level criteria. We accompany each criterion with a short and concise definition. The outline-level criteria we considered are: logical structure (the extent to which the generated outline follows a logical progression), depth (the extent to which the depth of the outline is appropriate for an NLP survey paper). The full survey-level criteria are: critical analysis (The depth of engagement with the surveyed literature beyond mere description), and synthesis (The ability to identify patterns, trends, and relationships across the body of literature).

For each criterion, we fed the definition to an LLM (\texttt{claude-sonnet-4-5-20250929}) and we asked it to generate two different prompts aimed at evaluating the same criteria. This step is aimed at testing the robustness of the results to prompt variation. We asked the LLM to provide prompts in chain-of-thought-style \cite{10.5555/3600270.3602070}, and to ask for scores from 1 to 5. 
%Automatic prompt generation minimizes human intervention by reducing the risk of arbitrariness and hacking in evaluation. 
For each criterion, we average the results across prompts and models. 

%Specifically, we evaluate generated surveys using multiple judge models, including both proprietary and open models, and assess complementary criteria at different levels of abstraction. Outline-level criteria measure \emph{logical structure} (coherence and progression of the outline) and \emph{depth} (appropriateness of outline granularity). Full-survey criteria evaluate \emph{critical analysis} (engagement beyond descriptive enumeration) and \emph{synthesis} (the ability to identify patterns, trends, and relationships across the literature). For each criterion, we automatically generate multiple evaluation prompts and aggregate scores across prompts and judge models to obtain more stable estimates.

\paragraph{Statistical testing.}
For each metric, we apply paired t-tests to assess the statistical significance of differences between \textsc{StructSurvey} and \textsc{SurveyForge}.

\section{Results}

\begin{table}[t]
\centering
\footnotesize
\setlength{\tabcolsep}{3pt} % tighter columns
\begin{tabular}{@{}lccc@{}}
\toprule
\textbf{Average} & \textbf{Gold} & \textbf{\textsc{SurveyForge}} & \textbf{\textsc{StructSurvey}} \\ \midrule
Words / Survey  & 6220.7 & 3954.5 & 4221.9 \\
1st-Level Sect. & 7.73   & 5.67   & 6.03   \\
Refs / Survey   & 109.0  & 49.8   & 49.8   \\
Subsect./Sect.  & 1.28   & 4.15   & 4.40   \\ \bottomrule
\end{tabular}
\caption{Comparison of structural metrics across systems.}
\label{tab:corpus_statistics}
\end{table}

\begin{table}[t]
\centering
\small % Use smaller font size to fit more content
\setlength{\tabcolsep}{3pt} % Reduce column separation
\begin{tabular}{lccc}
\toprule
\textbf{Metric} & \textbf{\textsc{SurveyForge}} & \textbf{\textsc{StructSurvey}} & \textbf{Sig.} \\
\midrule
\multicolumn{4}{l}{\textit{\textbf{ROUGE-1}}} \\
Prec. & $.6354 \pm .0461$ & $.6355 \pm .0378$ & ns \\
Recall & $.4194 \pm .0393$ & $\mathbf{.4480} \pm .0437$ & ** \\
F-meas. & $.5034 \pm .0313$ & $\mathbf{.5241} \pm .0342$ & *** \\
\midrule
\multicolumn{4}{l}{\textit{\textbf{ROUGE-2}}} \\
Prec. & $.1734 \pm .0233$ & $.1776 \pm .0227$ & ns \\
Recall & $.1148 \pm .0193$ & $\mathbf{.1252} \pm .0188$ & *** \\
F-meas. & $.1376 \pm .0196$ & $\mathbf{.1465} \pm .0193$ & *** \\
\bottomrule
\end{tabular}
\vspace{-2pt} % Reduce space after table
\caption*{
\small
\textbf{Sig.}: ns: not significant; **: $p<0.01$; ***: $p<0.001$. Best scores per metric are bolded.
}
\caption{ROUGE Performance Comparison (M$\pm$SD)}
\label{tab:rouge_results}
\end{table}

\begin{table}[t]
\centering
\small
\setlength{\tabcolsep}{2pt}
\begin{tabular}{lccc}
\toprule
\textbf{Criterion} & \textbf{\textsc{StructSurvey}} & \textbf{\textsc{SurveyForge}} & \textbf{p-value} \\
\midrule
Logical structure & 3.725 & 3.575 & 0.080 \\
Depth & 2.808 & 2.667 & 0.022* \\
Critical analysis & 3.025 & 3.042 & 0.783 \\
Synthesis & 3.767 & 3.650 & 0.032* \\
\bottomrule
\end{tabular}
\caption{LLM-as-a-Judge evaluation scores across different criteria. Asterisks (*) indicate statistical significance at $p < 0.05$.}
\label{tab:llm-judge-eval}
\end{table}

\subsection{Analysis of Generated Texts}

We analyze the structural and lexical properties of the generated surveys in comparison to gold human-written surveys. Table~\ref{tab:corpus_statistics} summarizes key statistics, and representative qualitative examples are provided in Appendix~\ref{sec:output_examples}. In the provided examples, we aligned each gold-standard section with the corresponding section in the outputs generated by the two systems. In the examples, we can see how the graph-generated outlines are more comprehensive and detailed, and how they align more often with the gold-standard content. Given that we imposed a limit of 500-800 words to each section draft, both automated systems produce surveys that are substantially shorter than human-authored ones — approximately one-third shorter on average.
%— highlighting the inherent difficulty of generating long, information-dense technical documents.

Despite this gap, \textsc{StructSurvey} consistently generates longer and more detailed surveys than \textsc{SurveyForge} (4{,}221.9 vs.\ 3{,}954.5 words on average). This difference suggests improved topical coverage and more effective content planning when structured retrieval is available.

We attribute this gain primarily to differences in outline construction. Although both systems are initialized with the same embedding-based retrieval results, their planners diverge in how they expand the outline. \textsc{StructSurvey} produces more fine-grained first-level outlines (6.03 sections on average, compared to 5.67 for \textsc{SurveyForge}). The availability of structured query functions encourages the planner to surface a broader range of themes and to differentiate them more clearly, better reflecting the organizational patterns observed in human-written surveys.

This effect is even more pronounced at the subsection level. As shown in Table~\ref{tab:corpus_statistics}, \textsc{StructSurvey} produces deeper outline decompositions, introducing a larger number of subsections per topic. While human surveys often favor more compact subsection structures, the additional granularity induced by \textsc{StructSurvey} provides more specific prompts for downstream content generation, which appears to translate into improved performance in subsequent evaluation. Overall, these findings suggest that structured retrieval not only expands access to relevant evidence but also actively shapes the conceptual organization of generated surveys.

\subsection{ROUGE Performance}

We evaluate both systems using ROUGE-1 and ROUGE-2 precision, recall, and F1, with results reported in Table~\ref{tab:rouge_results}. Precision is statistically indistinguishable across systems, indicating that \textsc{StructSurvey} does not introduce extraneous or off-topic content despite generating longer outputs.

In contrast, \textsc{StructSurvey} achieves consistent and statistically significant gains in recall. ROUGE-1 recall increases from 0.419 to 0.448 (+2.86 percentage points), while ROUGE-2 recall improves from 0.115 to 0.125 (+1.04 percentage points), yielding higher F1 scores for both metrics. Paired t-tests confirm that these gains in recall and F1 are statistically significant, with medium to large effect sizes.

To test whether ROUGE gains are driven only by longer outputs, we computed, for each survey, the ROUGE F1 difference ($\Delta$) between \textsc{StructSurvey} and \textsc{SurveyForge} for both ROUGE-1 and ROUGE-2, as well as the corresponding output-length $\Delta$. We then ranked both quantities and computed the Spearman correlation coefficient $\rho$ between the rankings. A value of $\rho=1$ would indicate perfect association between longer \textsc{StructSurvey} outputs and larger ROUGE gains. The analysis shows that length is a contributing factor, but not the only one: the correlation is higher for ROUGE-1 ($\rho=0.674$) and substantially lower for ROUGE-2 ($\rho=0.394$). This suggests that as evaluation moves from unigram overlap to higher-level lexical overlap, the contribution of length decreases.

%In order to understand whether the differences in ROUGE scores are only due to the graph-system outputs being longer, we performed an analysis where we computed for each example the F1-ROUGE $\Delta$ ($Score \Delta$) (the difference between the scores for \textsc{StructSurvey} and \textsc{SurveyForge}, for both ROUGE-1 and ROUGE-2), and the $Length \Delta$ (the difference between the lengths of the two system outputs). Then, we ranked both $Score \Delta$ and $Length \Delta$ values from smallest to highest and we computed the Spearman correlation coefficient $\rho$ between the two rankings. A $\rho$ of 1 shows that there is perfect correlation between the rankings (when the graph output is longer, then it also has better F1). The analysis shows that while output length is a contributing factor, it is not the only factor in determining the F1 of the two systems. While the correlation is higher ($\rho=0.674$) if we consider ROUGE-1, it is considerably lower ($\rho=0.394$) if we consider ROUGE-2. This suggests that, when we move from word overlap to more higher-level lexical overlap, the contribution of length becomes smaller. 

Overall, these results indicate that structured retrieval enables \textsc{StructSurvey} to cover a broader and more relevant set of concepts and terminology present in the reference surveys, improving topical coverage without sacrificing precision.

\paragraph{Added value of different graph query types.}
To identify which graph query types add the most value, we computed, for each survey, the ROUGE $\Delta$ between \textsc{StructSurvey} and \textsc{SurveyForge}, and correlated it with the frequency of each graph query function. The strongest correlation is observed for \texttt{related\_entity\_search} ($\rho=0.380$), despite its lower average usage than \texttt{single\_entity\_search} (3.1 vs.\ 7.1 uses per paper). By comparison, \texttt{single\_entity\_search} has correlation $0.197$, while \texttt{pair\_of\_related\_entities\_search} has correlation $0.114$. This suggests that starting from an entity and exploring its relationships is a particularly useful pattern for aligning generated surveys with human-written ones.

%\paragraph{Added value of different graph query types.}
%To identify which graph query types add the most value, we performed the following analyses: for each survey in our dataset, we calculated the Rouge $\Delta$, which is the difference between the Rouge score of StructSurvey and the SurveyForge one. Then, for each survey we computed the distribution of different graph query functions. Finally, we computed the Spearman correlation between the deltas and the frequency of occurrence of each function. The results show that the second pattern (\texttt{related\_entity\_search}) is the one with the highest correlation ($\rho=0.380$), despite it having much lower frequency of usage than the \texttt{single\_entity\_search} pattern (3.1 uses on average per paper vs. 7.1). The \texttt{single\_entity\_search} pattern has correlation $0.197$, while \texttt{pair\_of\_related\_entities\_search} has correlation $0.114$. This analysis suggests that starting from an entity and exploring its relations is a powerful search pattern that impacts the alignment of LLM-generated surveys with human ones.  
\subsection{LLM-as-a-Judge Evaluation}
\label{sec:llm_as_judge}

Table~\ref{tab:llm-judge-eval} reports LLM-as-a-Judge evaluation scores across the different criteria. Judge models consistently favor the outlines produced by \textsc{StructSurvey} in terms of both logical structure and depth, indicating clearer progression and more appropriate granularity. We also observe a statistically significant improvement in \emph{synthesis}, suggesting that structured retrieval enables the system to more effectively integrate information across multiple sources — one of the defining challenges of survey writing.

At the same time, the results highlight a persistent limitation shared by current LLM-based survey generation systems, including \textsc{StructSurvey}: limited \emph{critical analysis}. While \textsc{StructSurvey} remains competitive on this criterion, generated surveys still tend to emphasize descriptive coverage rather than critical evaluation of prior work.

\section{Conclusion}

We presented \textsc{StructSurvey}, a hierarchical multi-agent framework for automated survey paper generation that shifts structural reasoning from generation to retrieval through explicit extraction of entities, relations, and taxonomic groupings. By making structural information available during planning and retrieval, \textsc{StructSurvey} enables more coherent organization and synthesis of scientific literature.

Experiments on a new reference-grounded benchmark of 33 ACL survey papers (listed in Appendix~\ref{app:dataset_papers}) show that \textsc{StructSurvey} improves both lexical coverage and higher-level survey qualities — such as logical structure, depth, and synthesis — without sacrificing precision. These results highlight the value of structured retrieval for long-form scientific summarization and demonstrate that explicit structural context can meaningfully guide large language models in organizing complex research areas.

\section{Limitations}

\paragraph{Computational cost.}
A primary limitation of \textsc{StructSurvey} is its higher computational cost compared to embedding-only retrieval pipelines. While vector-based queries rely solely on embedding similarity search and incur no additional LLM calls beyond embedding computation—typically running in $O(\log D)$ time over an index of $D$ documents—structured queries depend on LLM-based information extraction and are substantially more expensive.

Let $E$ denote the number of extracted entities and $b$ the batch size used for entity classification. Executing a single structured query requires:
\begin{itemize}
    \item $D$ parallel LLM calls for entity extraction,
    \item one LLM call for category discovery,
    \item $\lceil E / b \rceil$ parallel LLM calls for entity classification,
\end{itemize}
resulting in a total LLM-call cost of
\[
    \mathrm{Cost}_{\mathrm{LLM}} = D + 1 + \left\lceil \frac{E}{b} \right\rceil .
\]

Structured queries produce local graph fragments that are merged into a global domain graph and reused across subsequent outline expansion and section drafting steps (Algorithm~\ref{alg:generate_survey}), which partially amortizes extraction cost. Nevertheless, overall cost still scales linearly with the number of structured queries and the choice of LLM model, making efficiency a key challenge for large-scale deployment.

\paragraph{Reference-grounded and abstract-only setting.}
Our evaluation is conducted in a reference-grounded setting, where the retrieval corpus for each survey consists exclusively of the papers cited by the gold survey, and retrieval operates at the abstract level rather than over full texts. This design isolates the core contribution of this work—structural planning, retrieval, and synthesis given relevant literature—but does not capture challenges related to open-corpus discovery, retrieval precision, or long-document processing. Extending structured retrieval to open and heterogeneous corpora is an important direction for future work.

\paragraph{Limited critical analysis.}
Despite improvements in structure, depth, and synthesis, generated surveys continue to exhibit limited critical analysis, often favoring descriptive aggregation over evaluative comparison or normative judgment. This limitation is shared by prior LLM-based survey-generation systems and reflects the difficulty of modeling expert-level critique and assessment of scientific trade-offs.

\paragraph{Evaluation limitations.}
Although we adopt a principled and robustness-oriented LLM-as-a-Judge evaluation protocol, automated evaluation of long-form scientific writing remains inherently imperfect. Metrics such as ROUGE capture lexical overlap but not conceptual correctness, while LLM-based judgments are sensitive to model capabilities and prompt design. Human expert evaluation would provide stronger validation but is costly and difficult to scale for long documents.

Together, these limitations point to promising avenues for future research, including more efficient structured retrieval, extension to open-corpus and full-text settings, stronger modeling of critical analysis, and deeper human-centered evaluation of generated surveys.

\section*{Acknowledgments}
We thank David Rosenberg for his detailed and insightful feedback, which substantially improved this paper.

\bibliography{custom}

\appendix
\section{Surveys Dataset Papers}
\label{app:dataset_papers}

This appendix lists the survey papers used to construct the reference-grounded evaluation dataset.
Entries are listed in alphabetical order by the surname of the first author.
The papers are documented for dataset reproducibility purposes and do not constitute additional paper references.

\begin{enumerate}%[leftmargin=*, label={[D\arabic*]}, itemsep=0.6ex]

\item Briakou, E., Agrawal, S., Zhang, K., Tetreault, J., and Carpuat, M. (2021).
\emph{A Review of Human Evaluation for Style Transfer}.
Proceedings of the First Workshop on Natural Language Generation, Evaluation, and Metrics (GEM), 58--67.

\item Chen, Y.-P., Nishida, N., Nakayama, H., and Matsumoto, Y. (2024).
\emph{Recent Trends in Personalized Dialogue Generation: A Review of Datasets, Methodologies, and Evaluations}.
Proceedings of LREC--COLING 2024, 13650--13665.

\item Chu, C., and Wang, R. (2018).
\emph{A Survey of Domain Adaptation for Neural Machine Translation}.
Proceedings of the 27th International Conference on Computational Linguistics, 1304--1319.

\item Cui, W., Yu, D., Jiao, X., Meng, Z., Zhang, G., Wang, Q., Guo, S.~Y., and King, I. (2025).
\emph{Recent Advances in Speech Language Models: A Survey}.
Proceedings of ACL 2025 (Long Papers), 13943--13970.

\item Dong, Q., Li, L., Dai, D., Zheng, C., Ma, J., Li, R., Xia, H., Xu, J., Wu, Z., Chang, B., Sun, X., Li, L., and Sui, Z. (2024).
\emph{A Survey on In-Context Learning}.
Proceedings of EMNLP 2024, 1107--1128.

\item Dong, Z., Zhou, Z., Yang, C., Shao, J., and Qiao, Y. (2024).
\emph{Attacks, Defenses and Evaluations for LLM Conversation Safety: A Survey}.
Proceedings of NAACL 2024, 6734--6747.

\item Geng, J., Cai, F., Wang, Y., Koeppl, H., Nakov, P., and Gurevych, I. (2024).
\emph{A Survey of Confidence Estimation and Calibration in Large Language Models}.
Proceedings of NAACL 2024, 6577--6595.

\item Iacob, R.~C.~A., Brad, F., Apostol, E.-S., Truică, C.-O., Hosu, I.~A., and Rebedea, T. (2020).
\emph{Neural Approaches for Natural Language Interfaces to Databases: A Survey}.
Proceedings of COLING 2020, 381--395.

\item Kim, M., Kim, M., Kim, H., Kwak, B.-W., Kang, S., Yu, Y., Yeo, J., and Lee, D. (2024).
\emph{Pearl: A Review-driven Persona-Knowledge Grounded Conversational Recommendation Dataset}.
Findings of ACL 2024, 1105--1120.

\item Lee, H.-Y., Li, S.-W., and Vu, T. (2022).
\emph{Meta Learning for Natural Language Processing: A Survey}.
Proceedings of NAACL 2022, 666--684.

\item Li, M., Zhao, Y., Zhang, W., Li, S., Xie, W., Ng, S.-K., Chua, T.-S., and Deng, Y. (2025).
\emph{Knowledge Boundary of Large Language Models: A Survey}.
Proceedings of ACL 2025 (Long Papers), 5131--5157.

\item Li, Z., Qu, L., and Haffari, G. (2020).
\emph{Context Dependent Semantic Parsing: A Survey}.
Proceedings of COLING 2020, 2509--2521.

\item Lu, P., Qiu, L., Yu, W., Welleck, S., and Chang, K.-W. (2023).
\emph{A Survey of Deep Learning for Mathematical Reasoning}.
Proceedings of ACL 2023 (Long Papers), 14605--14631.

\item Qin, L., Pan, W., Chen, Q., Liao, L., Yu, Z., Zhang, Y., Che, W., and Li, M. (2023).
\emph{End-to-end Task-oriented Dialogue: A Survey of Tasks, Methods, and Future Directions}.
Proceedings of EMNLP 2023, 5925--5941.

\item Ramisch, C., Walsh, A., Blanchard, T., and Taslimipoor, S. (2023).
\emph{A Survey of MWE Identification Experiments: The Devil is in the Details}.
Proceedings of the MWE Workshop 2023, 106--120.

\item Sadeddine, Z., Opitz, J., and Suchanek, F. (2024).
\emph{A Survey of Meaning Representations: From Theory to Practical Utility}.
Proceedings of NAACL 2024, 2877--2892.

\item Shen, X., Vakulenko, S., del Tredici, M., Barlacchi, G., Byrne, B., and de~Gispert, A. (2023).
\emph{Neural Ranking with Weak Supervision for Open-Domain Question Answering: A Survey}.
Findings of EACL 2023, 1736--1750.

\item Urlana, A., Mishra, P., Roy, T., and Mishra, R. (2024).
\emph{Controllable Text Summarization: Unraveling Challenges, Approaches, and Prospects}.
Findings of ACL 2024, 1603--1623.

\item Wang, L., Liu, S., Xu, M., Song, L., Shi, S., and Tu, Z. (2023).
\emph{A Survey on Zero Pronoun Translation}.
Proceedings of ACL 2023 (Long Papers), 3325--3339.

\item Wang, X., Wang, H., and Yang, D. (2022).
\emph{Measure and Improve Robustness in NLP Models: A Survey}.
Proceedings of NAACL 2022, 4569--4586.

\item Wang, Y., Wang, M., Manzoor, M.~A., Liu, F., Georgiev, G.~N., Das, R.~J., and Nakov, P. (2024).
\emph{Factuality of Large Language Models: A Survey}.
Proceedings of EMNLP 2024, 19519--19529.

\item Xia, P., Wu, S., and Van~Durme, B. (2020).
\emph{Which *BERT? A Survey Organizing Contextualized Encoders}.
Proceedings of EMNLP 2020, 7516--7533.

\item Xia, Z., Xu, J., Zhang, Y., and Liu, H. (2025).
\emph{A Survey of Uncertainty Estimation Methods on Large Language Models}.
Findings of ACL 2025, 21381--21396.

\item Yan, Y., Su, J., He, J., Fu, F., Zheng, X., Lyu, Y., Wang, K., Wang, S., Wen, Q., and Hu, X. (2025).
\emph{A Survey of Mathematical Reasoning in the Era of Multimodal Large Language Models: Benchmark, Method \& Challenges}.
Findings of ACL 2025, 11798--11827.

\item Ye, H., Zhang, N., Chen, H., and Chen, H. (2022).
\emph{Generative Knowledge Graph Construction: A Review}.
Proceedings of EMNLP 2022, 1--17.

\item Yu, X., Chatterjee, T., Asai, A., Hu, J, and Choi, E. (2022).
\emph{Beyond Counting Datasets: A Survey of Multilingual Dataset Construction and Necessary Resources}.
Findings of EMNLP 2022, 3725--3743.

\item Youssef, P., Koraş, O., Li, M., Schlötterer, J., and Seifert, C. (2023).
\emph{Give Me the Facts! A Survey on Factual Knowledge Probing in Pre-trained Language Models}.
Findings of EMNLP 2023, 15588--15605.

\item Zhang, Z., Fang, M., Chen, L., Namazi-Rad, M.-R., and Wang, J. (2023).
\emph{How Do Large Language Models Capture the Ever-changing World Knowledge? A Review of Recent Advances}.
Proceedings of EMNLP 2023, 8289--8311.

\item Zhang, Z., Yu, W., Yu, M., Guo, Z., and Jiang, M. (2023).
\emph{A Survey of Multi-task Learning in Natural Language Processing: Regarding Task Relatedness and Training Methods}.
Proceedings of EACL 2023, 943--956.

\item Zhang, Q., Chen, S., Xu, D., Cao, Q., Chen, X., Cohn, T., and Fang, M. (2023).
\emph{A Survey for Efficient Open Domain Question Answering}.
Proceedings of ACL 2023 (Long Papers), 14447--14465.

\item Zhao, R., Chen, H., Wang, W., Jiao, F., Do, X.~L., Qin, C., Ding, B., Guo, X., Li, M., Li, X., and Joty, S. (2023).
\emph{Retrieving Multimodal Information for Augmented Generation: A Survey}.
Findings of EMNLP 2023, 4736--4756.

\item Zhou, J., and Bhat, S. (2021).
\emph{Paraphrase Generation: A Survey of the State of the Art}.
Proceedings of EMNLP 2021, 5075--5086.

\item Zhou, Y., Ringeval, F., and Portet, F. (2023).
\emph{A Survey of Evaluation Methods of Generated Medical Textual Reports}.
Proceedings of the 5th Clinical Natural Language Processing Workshop, 447--459.

\end{enumerate}

\section{Prompts}

\subsubsection{Entity Extraction Prompt}
\label{sec:entity_extraction_prompt}

\textbf{System Prompt:}
\begin{lstlisting}[style=promptstyle]
  System Prompt:
  You are an expert at extracting specific entities from academic abstracts.

  Your task: Extract all instances of "{query}" from the given abstract.

  For each entity you find:
  1. Identify the specific name/type of the entity
  2. Describe its role or use in the abstract (1-2 sentences)

  Format your response EXACTLY as follows (one entity per block):

  ENTITY: [entity name]
  DESCRIPTION: [brief description of its role]

  ENTITY: [entity name]
  DESCRIPTION: [brief description of its role]

  If you find NO entities matching the query, respond with:
  NO_ENTITIES_FOUND

  Important:
  - Be specific about entity names
  - Only extract entities that clearly match the query type
  - Describe how the entity is used in THIS specific abstract

  User Message:
  Paper: {paper_title}

  Abstract:
  {abstract}

  Extract all instances of: {query}
\end{lstlisting}

\textbf{User Prompt:}
\begin{lstlisting}[style=promptstyle]

  Paper: {paper_title}

  Abstract:
  {abstract}

  Extract all instances of: {query}
\end{lstlisting}

\subsubsection{Category Discovery Prompt}
\label{sec:category_discovery_prompt}

\textbf{System Prompt:}
\begin{lstlisting}[style=promptstyle]
     You are an expert at categorizing entities based on specific criteria.

  You have been given a list of entities extracted from academic papers.

  Categorization Criterion: {categorize_by}

  Your task: Analyze the entities and propose 3-8 clear, mutually exclusive categories that cover most entities.

  Format your response EXACTLY as follows (one category per line):

  CATEGORY: [category name]
  CATEGORY: [category name]
  CATEGORY: [category name]
  ...

  Important:
  - Categories should be specific and meaningful
  - Categories should align with the categorization criterion
  - Aim for 3-8 categories (not too few, not too many)
  - Use clear, descriptive names
\end{lstlisting}

\textbf{User Prompt:}
\begin{lstlisting}[style=promptstyle]
     Here are the extracted entities:

  {entities_text}

  Total entities: {entity_count}

  Propose categories based on: {categorize_by}
\end{lstlisting}

\subsubsection{Entity Classification Prompt}
\label{sec:entity_classification_prompt}

\textbf{System Prompt}
\begin{lstlisting}[style=promptstyle]
You are an expert at classifying entities into predefined categories.

  Available categories:
  {categories_text}

  Your task: Classify each entity into ONE of the categories above.

  Format your response EXACTLY as follows (one classification per line):

  ENTITY: [entity name]
  CATEGORY: [category name]

  ENTITY: [entity name]
  CATEGORY: [category name]

  Important:
  - Each entity must be assigned to exactly ONE category
  - Use the exact category names from the list above
  - If an entity doesn't fit well, assign it to "Other"
  - Base classification on the entity's description
\end{lstlisting}

\textbf{User Prompt:}
\begin{lstlisting}[style=promptstyle]
      Classify these entities:

  {entities_text}

  Assign each entity to one of the categories listed above.
\end{lstlisting}

\subsubsection{Relationship-Based Entity Extraction}
\label{sec:relationship_based_entity_extraction_prompt}

\textbf{System Prompt:}
\begin{lstlisting}[style=promptstyle]
   You are an expert at extracting entity relationships from academic abstracts.

  Your task: Find all instances of "{find_entities}" that are related to "{fixed_entity}" in the given abstract.

  For each related entity you find:
  1. Identify the specific name/type of the entity
  2. Describe its relationship with "{fixed_entity}" in detail (2-3 sentences)
  3. Include the paper context to clarify where this information comes from

  Format your response EXACTLY as follows (one entity per block):

  ENTITY: [entity name]
  DESCRIPTION: In the paper "[paper_title]", [detailed description of how this entity relates to {fixed_entity}, including specific context and use case]

  ENTITY: [entity name]
  DESCRIPTION: In the paper "[paper_title]", [detailed description of the relationship]

  If you find NO entities matching the criteria, respond with:
  NO_ENTITIES_FOUND

  Important:
  - Be specific about entity names
  - Only extract entities that have a clear relationship with "{fixed_entity}"
  - Descriptions must be detailed and explain the relationship clearly
  - Always start descriptions with "In the paper [paper_title]," to provide context
  - Focus on HOW the entity relates to the fixed entity, not just what it is  
\end{lstlisting}

\textbf{User Prompt:}
\begin{lstlisting}[style=promptstyle]
  Paper: {paper_title}

  Abstract:
  {abstract}

  Fixed Entity: {fixed_entity}
  Find Entities: {find_entities}

  Extract all instances of "{find_entities}" that are related to "{fixed_entity}".

\end{lstlisting}

\subsubsection{Entity Pair Extraction Prompt}
\label{sec:entity_pair_extraction_prompt}
\textbf{System Prompt:}
\begin{lstlisting}[style=promptstyle]
 You are an expert at extracting entity relationships from academic abstracts.

  Your task: Find pairs of entities where:
  - Entity A is: "{entity_type_a}"
  - Entity B is: "{entity_type_b}"
  - There is a clear relationship or connection between them in the paper

  For each pair you find:
  1. Identify the specific name/type of Entity A
  2. Identify the specific name/type of Entity B
  3. Describe their relationship in detail (2-3 sentences)
  4. Include the paper context to clarify where this information comes from

  Format your response EXACTLY as follows (one pair per block):

  ENTITY_A: [entity A name]
  ENTITY_B: [entity B name]
  DESCRIPTION: In the paper "[paper_title]", [detailed description of how these entities relate, their connection, and the context of their use]

  ENTITY_A: [entity A name]
  ENTITY_B: [entity B name]
  DESCRIPTION: In the paper "[paper_title]", [detailed description of the relationship]

  If you find NO pairs matching the criteria, respond with:
  NO_PAIRS_FOUND

  Important:
  - Be specific about entity names
  - Only extract pairs that have a clear relationship
  - Descriptions must explain HOW the entities relate to each other
  - Always start descriptions with "In the paper [paper_title]," to provide context
  - Focus on the CONNECTION between the two entities
\end{lstlisting}

\textbf{User Prompt:}
\begin{lstlisting}[style=promptstyle]
      Paper: {paper_title}

  Abstract:
  {abstract}

  Entity Type A: {entity_type_a}
  Entity Type B: {entity_type_b}

  Extract all pairs where Entity A relates to Entity B.
\end{lstlisting}

\subsubsection{First-Level Outline Generation Prompt}
\label{sec:first_level_outline_generation_prompt}
\textbf{System Prompt (full version):}
\begin{lstlisting}[style=promptstyle]
 You are an expert survey paper writer. Your task is to create a first-level outline for a survey paper.

  Survey Title: {survey_title}
  Survey Abstract: {survey_abstract}
  Topic: {topic}

  Here are relevant papers from the literature to help you understand the field:

  {abstracts_text}

  Generate a first-level outline with 4-6 main sections. For each section:
  1. Provide a clear, descriptive title
  2. Choose a query type (vector or graph) that best suits the section's needs
  3. Generate the appropriate query based on the chosen type

  QUERY TYPES:
  You can generate two types of queries:

  1. VECTOR QUERY: A traditional semantic search query for retrieving relevant papers
     Format:
     TYPE: vector
     QUERY: [semantic search query text]

  2. GRAPH QUERY: A structured query for finding and categorizing entities/relationships
     Three patterns available:

     PATTERN 1 - Find and categorize entities of a given type:
     TYPE: graph_pattern_1
     QUERY: [entity type or property to find]
     CATEGORIZE_BY: [categorization criterion]

     PATTERN 2 - Find entities related to a fixed entity and categorize them:
     TYPE: graph_pattern_2
     FIXED_ENTITY: [specific entity name, e.g., "BERT model" or "question answering task"]
     FIND_ENTITIES: [type or property of related entities to find]
     CATEGORIZE_BY: [categorization criterion for found entities]

     PATTERN 3 - Find pairs of entity types and categorize both:
     TYPE: graph_pattern_3
     ENTITY_TYPE_A: [type or property of first entity set]
     ENTITY_TYPE_B: [type or property of second entity set]
     CATEGORIZE_A_BY: [categorization criterion for A entities]
     CATEGORIZE_B_BY: [categorization criterion for B entities]

  EXAMPLES:

  Example 1 - Vector Query:
  SECTION: Introduction to Neural Machine Translation
  TYPE: vector
  QUERY: neural machine translation architectures and approaches

  Example 2 - Graph Pattern 1:
  SECTION: Attention Mechanisms in Transformers
  TYPE: graph_pattern_1
  QUERY: attention mechanisms used in transformer models
  CATEGORIZE_BY: attention pattern type (self-attention, cross-attention, sparse attention, linear attention)

  Example 3 - Graph Pattern 2:
  SECTION: Applications of BERT
  TYPE: graph_pattern_2
  FIXED_ENTITY: BERT pre-trained model
  FIND_ENTITIES: downstream NLP tasks and applications
  CATEGORIZE_BY: task category (sequence classification, token classification, question answering, generation)

  Example 4 - Graph Pattern 3:
  SECTION: Evaluation Metrics and Model Architectures
  TYPE: graph_pattern_3
  ENTITY_TYPE_A: evaluation metrics for language generation
  ENTITY_TYPE_B: neural language model architectures
  CATEGORIZE_A_BY: aspect measured (fluency, coherence, factuality, diversity)
  CATEGORIZE_B_BY: architecture family (transformer-based, RNN-based, retrieval-augmented)

  Example 5 - Vector Query:
  SECTION: Challenges in Low-Resource Languages
  TYPE: vector
  QUERY: challenges and methods for NLP in low-resource languages

  Format your response EXACTLY as follows (each section on separate lines):

  SECTION: [Section Title]
  TYPE: [vector or graph_pattern_1 or graph_pattern_2 or graph_pattern_3]
  [Query fields based on type - see formats above]

  SECTION: [Next Section Title]
  TYPE: [type]
  [Query fields]
  ...

  Important: Start with an Introduction section and end with a Conclusion/Future Directions section.
\end{lstlisting}

\textbf{User Prompt:}
\begin{lstlisting}[style=promptstyle]
Generate the first-level outline for this survey on: {topic}
\end{lstlisting}

\subsubsection{Second-level Outline Generation Prompt}
\label{sec:second_level_outline_generation_prompt}
\textbf{System Prompt (Full Version):}
\begin{lstlisting}[style=promptstyle]

You are an expert survey paper writer. Your task is to create subsections for a specific section of a survey paper.

  Survey Title: {survey_title}
  Survey Topic: {topic}
  Section Title: {section_title}
  Section Focus: {section_query}

  Here is context from the literature for this section:

  {context_text}

  Generate 3-5 subsections for this section. For each subsection:
  1. Provide a clear, descriptive title that fits within the scope of the main section
  2. Choose a query type (vector or graph) that best suits the subsection's needs
  3. Generate the appropriate query based on the chosen type

  QUERY TYPES:
  You can generate two types of queries:

  1. VECTOR QUERY: A traditional semantic search query for retrieving relevant papers
     Format:
     TYPE: vector
     QUERY: [semantic search query text]

  2. GRAPH QUERY: A structured query for finding and categorizing entities/relationships
     Three patterns available:

     PATTERN 1 - Find and categorize entities of a given type:
     TYPE: graph_pattern_1
     QUERY: [entity type or property to find]
     CATEGORIZE_BY: [categorization criterion]

     PATTERN 2 - Find entities related to a fixed entity and categorize them:
     TYPE: graph_pattern_2
     FIXED_ENTITY: [specific entity name, e.g., "BERT model" or "question answering task"]
     FIND_ENTITIES: [type or property of related entities to find]
     CATEGORIZE_BY: [categorization criterion for found entities]

     PATTERN 3 - Find pairs of entity types and categorize both:
     TYPE: graph_pattern_3
     ENTITY_TYPE_A: [type or property of first entity set]
     ENTITY_TYPE_B: [type or property of second entity set]
     CATEGORIZE_A_BY: [categorization criterion for A entities]
     CATEGORIZE_B_BY: [categorization criterion for B entities]

  EXAMPLES:

  Example 1 - Vector Query:
  SECTION: Core Concepts and Terminology
  TYPE: vector
  QUERY: fundamental concepts and terminology in the field

  Example 2 - Graph Pattern 1:
  SECTION: Model Architectures
  TYPE: graph_pattern_1
  QUERY: neural network architectures for this task
  CATEGORIZE_BY: architectural design (encoder-only, decoder-only, encoder-decoder, hybrid)

  Example 3 - Graph Pattern 2:
  SECTION: BERT Variants and Extensions
  TYPE: graph_pattern_2
  FIXED_ENTITY: BERT model
  FIND_ENTITIES: model variants and extensions
  CATEGORIZE_BY: modification type (domain-specific, multilingual, efficient, task-specific)

  Example 4 - Graph Pattern 3:
  SECTION: Training Techniques and Datasets
  TYPE: graph_pattern_3
  ENTITY_TYPE_A: training techniques and optimization methods
  ENTITY_TYPE_B: benchmark datasets
  CATEGORIZE_A_BY: technique category (supervised, self-supervised, semi-supervised, reinforcement learning)
  CATEGORIZE_B_BY: dataset scale and domain (small-scale academic, large-scale web, domain-specific)

  Example 5 - Vector Query:
  SECTION: Recent Developments
  TYPE: vector
  QUERY: recent advances and novel approaches in the subfield

  Format your response EXACTLY as follows (each subsection on separate lines):

  SECTION: [Subsection Title]
  TYPE: [vector or graph_pattern_1 or graph_pattern_2 or graph_pattern_3]
  [Query fields based on type - see formats above]

  SECTION: [Next Subsection Title]
  TYPE: [type]
  [Query fields]
  ...

  Important: The subsections should logically break down the main section topic.

\end{lstlisting}

\textbf{User Prompt:}
\begin{lstlisting}[style=promptstyle]
Generate subsections for section: {section_title}
\end{lstlisting}

\subsubsection{Query Decomposition Prompt}
\label{sec:query_decomposition_prompt}

\textbf{System Prompt (Full Version):}
\begin{lstlisting}[style=promptstyle]

  You are an expert at decomposing search queries. Your task is to create specific sub-queries for retrieving relevant papers.

  Section: {section_title}
  Main Query: {section_query}

  Subsections:
  {subsections_text}

  Generate 3-5 specific search queries that would help retrieve relevant papers for this section and its subsections.
  Each query should be focused and specific. You can choose between vector and graph query types.

  QUERY TYPES:
  You can generate two types of queries:

  1. VECTOR QUERY: A traditional semantic search query for retrieving relevant papers
     Format:
     TYPE: vector
     QUERY: [semantic search query text]

  2. GRAPH QUERY: A structured query for finding and categorizing entities/relationships
     Three patterns available:

     PATTERN 1 - Find and categorize entities of a given type:
     TYPE: graph_pattern_1
     QUERY: [entity type or property to find]
     CATEGORIZE_BY: [categorization criterion]

     PATTERN 2 - Find entities related to a fixed entity and categorize them:
     TYPE: graph_pattern_2
     FIXED_ENTITY: [specific entity name, e.g., "BERT model" or "question answering task"]
     FIND_ENTITIES: [type or property of related entities to find]
     CATEGORIZE_BY: [categorization criterion for found entities]

     PATTERN 3 - Find pairs of entity types and categorize both:
     TYPE: graph_pattern_3
     ENTITY_TYPE_A: [type or property of first entity set]
     ENTITY_TYPE_B: [type or property of second entity set]
     CATEGORIZE_A_BY: [categorization criterion for A entities]
     CATEGORIZE_B_BY: [categorization criterion for B entities]

  EXAMPLES:

  Example 1 - Vector Query:
  TYPE: vector
  QUERY: recent advances in attention mechanisms for transformers

  Example 2 - Graph Pattern 1:
  TYPE: graph_pattern_1
  QUERY: pre-training objectives for language models
  CATEGORIZE_BY: objective type (masked language modeling, next sentence prediction, causal language modeling, denoising)

  Example 3 - Graph Pattern 2:
  TYPE: graph_pattern_2
  FIXED_ENTITY: GPT-3 model
  FIND_ENTITIES: applications and use cases
  CATEGORIZE_BY: application domain (text generation, code generation, question answering, translation)

  Example 4 - Graph Pattern 3:
  TYPE: graph_pattern_3
  ENTITY_TYPE_A: loss functions
  ENTITY_TYPE_B: model architectures
  CATEGORIZE_A_BY: loss type (cross-entropy, contrastive, ranking, generative)
  CATEGORIZE_B_BY: model size (small <100M params, medium 100M-1B, large >1B)

  Format your response as follows:

  TYPE: [vector or graph_pattern_1 or graph_pattern_2 or graph_pattern_3]
  [Query fields based on type]

  TYPE: [type]
  [Query fields]
  ...
\end{lstlisting}

\textbf{User Prompt:}
\begin{lstlisting}[style=promptstyle]
Generate sub-queries for section: {section_title}
\end{lstlisting}

\subsubsection{Section Writing Prompt}
\label{sec:section_writing_prompt}
\textbf{System Prompt:}
\begin{lstlisting}[style=promptstyle]
 You are an expert survey paper writer. Your task is to write a comprehensive section for a survey paper.

  Survey Title: {survey_title}
  Survey Topic: {topic}
  Section Title: {section_title}
  Section Focus: {section_query}

  Subsections to cover:
  {subsections_text}

  Here are relevant papers from the literature to reference:

  {docs_text}

  {graph_context}

  Write a comprehensive section that:
  1. Covers all subsections naturally (don't use subsection titles as headers)
  2. Synthesizes information from multiple papers
  3. If structured entity analysis is provided above, USE IT to organize and structure your discussion
     - Mention the categories and how they relate to each other
     - Use the entity counts to show prevalence of different approaches
     - Reference specific entities when discussing techniques
  4. Provides clear explanations and comparisons
  5. Uses proper academic writing style
  6. Is approximately 500-800 words
  7. IMPORTANT: Cite papers by their ACTUAL TITLES when referencing them
     - Good examples:
       * "The work 'Retrieval-Augmented Generation' introduces..."
       * "As shown in 'BERT: Pre-training of Deep Bidirectional Transformers', ..."
       * "Recent work on multimodal learning ('CLIP', 'DALL-E') has demonstrated..."
     - BAD examples (DO NOT USE):
       * "Paper 1 shows..."
       * "The first paper introduces..."
       * "As mentioned in [1]..."
  8. When multiple papers address the same topic, mention their titles together
  9. Use the paper titles as natural references in your writing

  CRITICAL FORMATTING INSTRUCTIONS:
  - DO NOT include the section title as a heading (e.g., "## {section_title}") in your output
  - Start directly with the section content
  - The section title will be added automatically by the system
  - You MUST use the actual paper titles in your citations, not generic labels like "Paper 1" or numbered references
  - If structured analysis is provided, integrate it naturally into your writing to organize the discussion
\end{lstlisting}

\textbf{User Prompt:}
\begin{lstlisting}[style=promptstyle]

Write the section: {section_title}

\end{lstlisting}

%\clearpage
\section{Examples of Outputs from the Systems}
\label{sec:output_examples}

\begin{table*}[t]
\centering
\small
\setlength{\tabcolsep}{4pt}
\begin{tabular}{p{0.3\textwidth}|p{0.3\textwidth}|p{0.3\textwidth}}
\toprule
\textbf{Gold Standard} & \textsc{SurveyForge} & \textsc{StructSurvey} \\
\midrule
\multicolumn{3}{c}{\textbf{Example: ``Factuality of Large Language Models: A Survey''}} \\
\midrule
\textbf{1 Introduction} & \textbf{1 Introduction} & \textbf{1 Introduction and Problem Definition} \\
& $\circ$ Characterizing Factuality and Hallucinations & $\circ$ Rise of LLMs and Imperative of Factuality \\
& $\circ$ Importance of Factuality in Applications & $\circ$ Defining and Characterizing Hallucinations \\
& $\circ$ Overview of the Factuality Problem & $\circ$ Scope and Significance of Research \\
\addlinespace
\textbf{2 Background} & \textbf{2 Understanding Factuality Issues} & \textbf{3 Causes and Contributing Factors} \\
$\circ$ Hallucination vs. Factuality & $\circ$ Data and Training Artifacts & $\circ$ Knowledge Gaps and Incomplete Understanding \\
$\circ$ Trustworthiness/Reliability vs. Factuality & $\circ$ Knowledge Representation and Retrieval Deficiencies & $\circ$ Biases and Distortions in Training Data \\
& $\circ$ Outdated Information and Temporal Drift & $\circ$ Model Architecture and Training Objectives \\
\addlinespace
\textbf{3 Evaluating Factuality} & \textbf{3 Evaluating LLM Factuality} & \textbf{2 Understanding and Quantifying Factuality} \\
$\circ$ 3.1 Datasets and Metrics & $\circ$ Benchmarking LLM Factuality & $\circ$ Taxonomy of Factuality Errors \\
$\circ$ 3.2 Other Metrics & $\circ$ Human vs. Automated Assessment & $\circ$ Automated Metrics and Frameworks \\
& $\circ$ Fine-grained Evaluation Methods & \textbf{5 Benchmarking and Evaluation} \\
& & $\circ$ Domain-Specific and Specialized Benchmarks \\
\addlinespace
\textbf{4 Improving Factuality} & \textbf{4 Mitigating LLM Factuality Issues} & \textbf{4 Mitigation Techniques} \\
$\circ$ 4.1 Pre-training & $\circ$ Retrieval-Augmented Generation & $\circ$ Retrieval-Augmented Generation (RAG) \\
$\circ$ 4.2 Tuning and RLHF & $\circ$ Fine-tuning and Knowledge Editing & $\circ$ Fine-tuning and Preference Optimization \\
\quad - Retrieval Augmentation & $\circ$ Verification, Rectification, and Debating & $\circ$ Prompting and Generation Strategies \\
$\circ$ 4.3 Inference & $\circ$ Context and Citation Enhancement & $\circ$ Model Editing and Multi-Agent Verification \\
\quad - 4.3.1 Decoding Strategy & & \\
\quad - 4.3.2 ICL and Self-reasoning & & \\
$\circ$ 4.4 Automatic Fact Checkers & & \\
\addlinespace
\textbf{5 Factuality of Multimodal LLMs} & \textbf{5 Challenges and Future Directions} & \textbf{6 Challenges in Automated Evaluation} \\
$\circ$ Existence, Attribute, and Relationship Factuality & $\circ$ Obstacles to Automated Fact-Checking & $\circ$ Ambiguity and Subjectivity in Assessment \\
$\circ$ Evaluation and Mitigation & $\circ$ Domain-Specific Factuality in LLMs & $\circ$ Atomic vs. Holistic Factuality \\
& $\circ$ Real-Time Verification & $\circ$ Dependence on External Knowledge \\
\addlinespace
\textbf{6 Challenges and Future Directions} & \textbf{6 Conclusion} & \textbf{7 Future Directions and Open Challenges} \\
$\circ$ Learning Distributions vs. Facts & $\circ$ Current State of Research & $\circ$ Dynamic Knowledge Updates \\
$\circ$ RAG Bottlenecks (Latency/Multi-hop) & $\circ$ Key Challenges and Limitations & $\circ$ Bridging Fluency and Factual Accuracy \\
$\circ$ Improving Retrieval and Fact-checker Accuracy & $\circ$ Role of Benchmarking & $\circ$ Hallucination Snowball Effects \\
\addlinespace
\textbf{7 Conclusion} & & \\
\bottomrule
\end{tabular}
\caption{Comparison of survey outlines for ``Factuality of Large Language Models: A Survey''. \textsc{StructSurvey} introduces a taxonomy of factuality errors, distinguishes domain-specific and specialized benchmarks, and covers a broader set of mitigation techniques and future directions than \textsc{SurveyForge}.}
\label{tab:factuality_comparison}
\end{table*}

\begin{table*}[t]
\centering
\small
\setlength{\tabcolsep}{4pt}
\begin{tabular}{p{0.3\textwidth}|p{0.3\textwidth}|p{0.3\textwidth}}
\toprule
\textbf{Gold Standard} & \textsc{SurveyForge} & \textsc{StructSurvey} \\
\midrule
\multicolumn{3}{c}{\textbf{Example: ``Context Dependent Semantic Parsing: A Survey''}} \\
\midrule
\textbf{1 Introduction} & \textbf{1 Introduction} & \textbf{1 Introduction} \\
$\circ$ CISP vs. CDSP & $\circ$ Rise of Context-Dependent Parsing & $\circ$ Evolution of Semantic Parsing \\
$\circ$ Challenges (Reference, Ellipsis) & $\circ$ Challenges in CDSP & $\circ$ Motivations for Context Dependence \\
\addlinespace
\textbf{2 Background} & \textbf{2 Foundations of Semantic Parsing} & \textbf{2 Foundational Concepts} \\
$\circ$ Meaning Representations (MRs) & $\circ$ Core Task and MRs & $\circ$ Knowledge Representation \\
$\circ$ Symbolic vs. Neural Approaches & $\circ$ Traditional vs. Modern Approaches & $\circ$ Algorithmic Approaches \\
\quad - SEQ2SEQ, SEQ2TREE, RNNs & & \\
$\circ$ \textbf{2.2 Evaluation} & \textbf{5 Evaluation and Benchmarks} & \textbf{5.4 Evaluation Metrics} \\
\quad - Exact/Set Match Accuracy & $\circ$ Semantic Parsing Metrics & $\circ$ Benchmarking and Frameworks \\
\addlinespace
\textbf{3 Context Dependent Semantic Parsing} & \textbf{3 CDSP Methods} & \textbf{3 Contextual Information} \\
$\circ$ Definition of Local Context & $\circ$ Dialogue History Awareness & $\circ$ Dialogue and Document Context \\
$\circ$ \textbf{3.1 Symbolic Approaches} & $\circ$ Temporal and Sequential Context & $\circ$ User-Specific and External Knowledge \\
$\circ$ \textbf{3.2 Neural Approaches} & $\circ$ Interactive and Iterative Parsing & \textbf{4 Methods for CDSP} \\
\quad - Context-aware Encoders/Decoders & & $\circ$ Encoder-Decoder Architectures \\
\quad - Copy Mechanisms \& Pointers & & $\circ$ Pointer Networks and Copying \\
$\circ$ \textbf{3.3 Neural-Symbolic Approaches} & & $\circ$ Structure-Aware Models \\
\addlinespace
\textbf{4 Datasets and Resources} & \textbf{4 Datasets and Tasks} & \textbf{5 Datasets and Evaluation} \\
$\circ$ 4.1 Scenarios (Single vs. Multi-party) & $\circ$ Cross-Domain and Text-to-SQL & $\circ$ Overview of CDSP Datasets \\
$\circ$ 4.2 Context and Annotations & $\circ$ Temporal and Interactive Datasets & $\circ$ Conversational SQL Datasets \\
& & $\circ$ Semantic Parse Correction \\
\addlinespace
\textbf{5 Challenges and Future Directions} & \textbf{6 Challenges and Future Directions} & \textbf{7 Challenges and Future Directions} \\
$\circ$ Far-side Pragmatics/Implicatures & $\circ$ Ambiguity and Complexity of Context & $\circ$ Data Scarcity and Generalization \\
$\circ$ Causal Structure Discovery & $\circ$ Role of Interaction and Feedback & $\circ$ Ambiguity and Knowledge Integration \\
$\circ$ Low-resource CDSP & $\circ$ Limited Supervision & $\circ$ Multimodal and Situated Context \\
\addlinespace
& & \textbf{6 Applications and Case Studies} \\
& & $\circ$ Text-to-SQL and Code Generation \\
& & $\circ$ Conversational AI and QA \\
\bottomrule
\end{tabular}
\caption{Comparison of survey outlines for ``Context Dependent Semantic Parsing: A Survey''. The graph-based method captures context types, method categories such as encoder--decoder architectures, pointer mechanisms, and structure-aware models, and includes data scarcity among future challenges.}
\label{tab:context_parsing_comparison}
\end{table*}

\begin{table*}[t]
\centering
\small
\setlength{\tabcolsep}{4pt}
\begin{tabular}{p{0.3\textwidth}|p{0.3\textwidth}|p{0.3\textwidth}}
\toprule
\textbf{Gold Standard} & \textsc{SurveyForge} & \textsc{StructSurvey} \\
\midrule
\multicolumn{3}{c}{\textbf{Example: ``Paraphrase Generation: A Survey of the State of the Art''}} \\
\midrule
\textbf{1 Introduction} & \textbf{1 Introduction} & \textbf{1 Introduction} \\
& $\circ$ Motivation and Significance & $\circ$ Defining Paraphrase Generation \\
& $\circ$ Evolution of Techniques & $\circ$ Historical Evolution \\
\addlinespace
\textbf{4 Traditional Approaches} & \textbf{2 Traditional and Rule-Based} & \textbf{2 Traditional and Rule-Based} \\
$\circ$ Rule-Based Approaches & $\circ$ Lexical Substitution & $\circ$ Word Reordering and Substitution \\
$\circ$ Thesaurus-Based Approaches & $\circ$ Syntactic Transformation & $\circ$ Template-Based Systems \\
$\circ$ SMT-Based Approach & $\circ$ Template-Based Paraphrasing & $\circ$ Early Data-Driven Approaches \\
\addlinespace
\textbf{5 Neural Approaches} & \textbf{3 Neural Network Models} & \textbf{3 Neural Paraphrase Models} \\
$\circ$ 5.1 Encoder-Decoder Architecture & $\circ$ Sequence-to-Sequence Models & $\circ$ Encoder-Decoder Architectures \\
\quad - Encoding Side & $\circ$ Transformer-based Generation & $\circ$ Attention and Transformer Models \\
\quad - Decoding Side & $\circ$ Adversarial and RL Approaches & $\circ$ RNNs and Specialized Architectures \\
\addlinespace
\textbf{5.2 Improvements (Advanced)} & \textbf{4 Advanced Techniques} & \textbf{4 Advanced Enhancements} \\
$\circ$ A. Model-focused (Attention, VAE) & $\circ$ Diverse Paraphrase Generation & $\circ$ Latent Representations for Diversity \\
$\circ$ B. Attribute-focused (Syntax, Multi) & $\circ$ Controlled Paraphrase Generation & $\circ$ Controlling Linguistic Properties \\
\quad - Explicit vs. Implicit Control & $\circ$ Unsupervised and Weakly Supervised & $\circ$ Adversarial and RL Improvements \\
\addlinespace
\textbf{2 Datasets / 3 Evaluation} & \textbf{4.5 Evaluation Metrics} & \textbf{5 Datasets, Eval, and Apps} \\
$\circ$ 2 Datasets (PPDB, MSCOCO, etc.) & $\circ$ Challenges in Generation Evaluation & $\circ$ Overview of PG Datasets \\
$\circ$ 3 Evaluation (BLEU, METEOR, ROUGE) & & $\circ$ Common Evaluation Metrics \\
& & $\circ$ Relationship: Datasets vs. Metrics \\
\addlinespace
\textbf{6 SOTA / 7 Conclusion} & \textbf{5 Applications / 6 Conclusion} & \textbf{6 Conclusion} \\
$\circ$ Pretrained Language Models & $\circ$ QA and Dialogue Systems & $\circ$ Challenges and Open Problems \\
$\circ$ Controllable Generation & $\circ$ Data Augmentation & $\circ$ Emerging Trends \\
$\circ$ Stylistic Paraphrasing & $\circ$ Bridging the Unsupervised Gap & $\circ$ Unsupervised and Cross-Lingual \\
\bottomrule
\end{tabular}
\caption{Comparison of survey outlines for ``Paraphrase Generation: A Survey of the State of the Art''. The graph-based output includes a dedicated paraphrase-generation dataset section aligned with the gold survey and distinguishes it from evaluation.}
\label{tab:paraphrase_survey_comparison}
\end{table*}

\end{document}